\documentstyle[12pt,epsfig]{article}
\begin{document}
\title{\quad  
{\Large{\bf New Analysis Strategy to Search for\\
            New Particles at LHC}}
\author{ Yongsheng Gao\\
         Physics Department\\
         Southern Methodist University\\
         Dallas, TX 75275-0175, USA\\
         Email: gao@mail.physics.smu.edu  } }
\maketitle

\begin{abstract}

The ATLAS and CMS experiments at LHC have great physics potential in 
discovering many possible new particles, from Standard Model (SM) 
Higgs boson to supersymmetric (SUSY) and other beyond the SM new particles
over a very large mass range around the TeV scale.
We first review the current analysis strategy in searching for new 
particles at LHC and its vulnerability to statistical fluctuations.
We then discuss an alternative new analysis strategy in order to avoid
this vulnerability. We use maximum likelihood method as an example and 
discuss the new challenges in using maximum likelihood method as such a 
new analysis strategy to search for new  particles at LHC.

\end{abstract}


\section{Introduction}
The Large Hadron Collider (LHC) at CERN near Geneva, Switzerland will 
open a new frontier in particle physics due to its higher collision energy 
and luminosity compared to the existing accelerators. 
The general-purpose ATLAS and CMS experiments at LHC will employ precision 
tracking, calorimetry and muon measurements over a large solid angle to 
accurately identify and measure electrons, muons, photons, jets and missing 
energy etc.
The guiding principle in designing the ATLAS and CMS detectors has been to 
maximize the discovery potential for new physics such as the Standard Model 
(SM) Higgs boson, supersymmetric (SUSY) and other beyond the SM new particles 
over a very large mass range around the TeV scale, while keep the capability 
of high-accuracy measurements of the known heavy quarks and gauge bosons.

With the start of the ATLAS and CMS data taking around 2007, it will 
be almost certain that we will observe some ``anomaly'' from SM predictions 
with the first ATLAS and CMS data. The first question we need to answer is: 
is the observed ``anomaly'' a true sign of new physics or simply due to 
statistical fluctuations?

In this paper we first review the current analysis strategy in searching for
new particles at LHC and its vulnerability to statistical fluctuations. 
We then discuss an alternative new analysis strategy in order to avoid this 
vulnerability. We use maximum likelihood method as an example and discuss the 
new challenges in using maximum likelihood method as such a new analysis 
strategy to search for new  particles at LHC.

The reliability and uncertainties in SM predictions for specific new particle 
searches due to physics, Monte Carlo modeling and simulation or detector 
performance are not in the scope of this paper.

\section{Current Analysis Strategy in Searching for New Particles at LHC}
The LHC experiments have been developing an extensive physics program to 
search for all the possible new particles accessible at LHC.
The current analysis strategy to determine the expected sensitivity for 
searching a new particle with specific mass and production/decay mode can 
in general be briefly described as follows:

  \begin{itemize}
       
     \vskip 0.1cm

     \item {Choose a specific production and decay mode of a certain new 
            particle with a specific mass (i.e. Signal). 
            Consider all the other possible processes (Background) that can 
            contribute to the same or similar final state of the Signal 
            process after the corresponding trigger requirements.}

     \vskip 0.2cm

     \item {Use Monte Carlo simulation to generate events of both the Signal 
            and Background processes with realistic detector responses, as well 
            as sufficient statistics according to the corresponding luminosity, 
            cross-sections and branching fractions etc.}

     \vskip 0.2cm

     \item {Study all the possible experimental observables that can provide 
            some discrimination power between the Signal and Background 
            processes. Apply cuts on these observables to suppress Background 
            events while keep Signal events as efficient as possible, in order 
            to be able to observe a peak or excess of Signal events over 
            Background events in the expected region of the invariant mass or 
            similar kinematic observable.}

     \vskip 0.2cm

     \item {The final event selection criteria is determined by optimizing the 
            number of Signal events ($S$) in the expected region of the invariant 
            mass or similar kinematic observable over the square root of the
            total number of Background events ($\sqrt{B}$) in the the same 
            region.
            The expected significance or sensitivity of the search for this new 
            particle with this specific mass and production/decay mode is given 
            by $S/ \sqrt{B}$ from these simulation study.}

  \end{itemize}

In order to obtain more discrimination power from the same experimental 
observables, more sophisticated technics (Artificial Neural Net (NN) or 
Likelihood) using these discriminative observables, instead of applying
simple cuts on them, are often used to achieve better $S/ \sqrt{B}$.

\section{Where is the Problem?}
The above analysis strategy is to search for statistically significant peak or 
excess of events in the expected region of the invariant mass or similar 
kinematic observable. The expected significance of the search is given by  
$S/ \sqrt{B}$ from simulation study described above. 
However, there exist some problems in this approach.

Because we do not know the mass of any of the new particles accessible at LHC, 
the real new physics signal from future ATLAS and CMS data can show up as a 
peak or excess of events anywhere 
in the entire range of the invariant mass or similar kinematic observable. 
Therefore, our current analysis strategy is to search for 
any statistically significant peak or excess of events in the entire range 
of the invariant mass or similar kinematic observable accessible at LHC.
Image that we do observe some peak or excess of events from the first ATLAS and 
CMS data. In order to rule out, beyond reasonable doubt, the possibility that 
the observed peak or excess of events is from statistical fluctuations,
we need to rule out the statistical fluctuation probability that such peak or 
excess of events can show up anywhere in the entire range of the invariant 
mass or similar kinematic observable we are searching. 
Apparently, the statistical fluctuation probability in this case is much   
higher than that of a peak or excess of events showing up exactly where we 
expect it for a new particle with specific mass as described in the previous 
section.

Divide up the search region into many smaller regions and search separately
does not help. While the statistical fluctuation probability of observing
a peak or excess of events is lower in a smaller range of the invariant mass or 
similar kinematic observable, it also increases as we search through multiple
such smaller ranges.

Therefore, the true significance of our observing a peak or excess of 
events in the invariant mass or similar kinematic observable from the 
incoming ATLAS and CMS data will be much lower than the expected significance 
determined by $S/ \sqrt{B}$ described in the previous section, 
even if we observe from data exactly the same peak or excess of events 
we are expecting from the simulation study. Our current analysis strategy 
to search for new particles is vulnerable to statistical fluctuations and 
will result in overestimation of the significance of our experimental 
observation from the future ATLAS and CMS data.

\section{Alternative New Analysis Strategy in Searching for New Particles 
         at LHC}
The additional statistical fluctuation probability described above, which 
comes from the fact that we do not know the mass of the new particles we
are searching, is not considered in the current analysis strategy to
search for new particles at LHC.
If this additional statistical fluctuation probability is taken into 
account, our current analysis strategy will result in reduced 
significance or sensitivity to new physics. 
In order to make full use of the discovery potential of the incoming ATLAS 
and CMS data, we need to think about new analysis strategy to search for
new particles at LHC.
In the following, we explore an alternative new strategy in order to 
avoid the above described vulnerability to statistical fluctuations while 
make full use of the discovery potential of the incoming ATLAS and CMS data.

The vulnerability to statistical fluctuations of our current analysis
strategy comes from the fact that we search for statistically significant 
peak or excess of events over a wide range of the invariant mass or 
similar kinematic observable. We need to change this in order to avoid
this vulnerability. In the mean time, the new analysis strategy needs to
be as efficient and sensitive as possible to the new physics we are searching.

Instead of searching for statistically significant peak or excess of events 
over a wide range of the invariant mass or similar kinematic observable,
an alternative new analysis strategy is to scan all the possible 
physics hypotheses allowed in the data sample and use all the information 
of the data to determine which physics hypotheses is favored by the data. 
Once the favored physics hypotheses is found, their significances can be 
determined from how far away statistically they are from any other possible 
physics hypotheses.
We use maximum likelihood method as an example and discuss the new challenges 
in using maximum likelihood method as such an alternative new analysis 
strategy to search for new  particles at LHC.

\section{Maximum Likelihood Method as an Alternative New Strategy in 
         Searching for New Particles at LHC}

Maximum likelihood method has been widely used in many high energy physics
(HEP) experiments. The unbinned maximum likelihood fit to the data sample
includes all possible physics hypotheses (Signal and Background processes). 
The likelihood of an event in the data sample is parameterized by the sum of 
probabilities for all relevant Signal and Background hypotheses, with relative 
weights to be determined by maximizing the likelihood function.
The probability of a particular hypothesis is calculated as a product 
of the probability density functions (PDFs) for each of the input experimental
(discriminative) observable. Some simple examples of using unbinned maximum 
likelihood method in HEP data analyses can be found in~\cite{bigrare,cleopv}.
Because no very tight cut on the discriminative observable is absolutely 
crucial, maximum likelihood method is known for being able to identify a 
handful of Signal events from huge Background events, as long as the Signal 
and the Background processes can be separated reasonably well by using all 
the experimental discriminative observables. i.e. it has the characteristic 
of being sensitive and efficient for Signal while insensitive to Background.

While maximum likelihood method has been widely used in HEP experiments, there 
are new challenges in using maximum likelihood method as an alternative new 
strategy in searching for new particles at LHC.

First, the scan of the physics hypotheses needs to cover all possible 
physics hypotheses allowed in the data sample. This means the Signal or 
Background processes, in many cases both, in the maximum likelihood fit can 
vary. For example, the SM Higgs and SUSY particles' masses are unknown. 
In many SUSY particle searches, the most serious background processes for 
the search come from the other possible SUSY particles. Therefore the possible 
new particles' masses need to be parameters of the maximum likelihood fit. 

Second, the correlations among all the discriminative observables in the 
maximum likelihood fit need to be taken into account properly. To maximize 
our sensitivity to search for new particles, we need to take full advantage 
of all the useful experimental observables that can provide us with some
discrimination power between the Signal and Background processes. 
Correlations often exist among these observables and they need to be taken 
into account correctly for the maximum likelihood method to give us reliable 
results.

Therefore, using unbinned maximum likelihood method as an alternative new
strategy to search for new particle at LHC is to scan all the allowed 
parameter(s) space  and perform unbinned maximum likelihood fit for each set 
of the allowed parameter(s) of all possible physics hypotheses in the data
sample. The procedure can be briefly described as follows:

  \begin{itemize}
       
     \vskip 0.1cm

     \item  {Choose a specific production and decay mode of a certain new 
            particle (i.e. Signal). Notice the mass of the new particle is not
            specified yet. Consider all the other possible processes 
            (Background) that can contribute to the same or similar final state 
            of the Signal process after the corresponding trigger requirements. 
            Again, the Background processes may also vary.}

     \vskip 0.2cm

     \item {Use Monte Carlo simulation to generate events of both the Signal 
            and Background processes with realistic detector responses, as well 
            as sufficient statistics according to the corresponding luminosity, 
            cross-sections and branching fractions etc. Large simulation samples
            are needed for each set of the allowed parameter(s) (i.e. new 
            particle' mass) for all the possible physics hypotheses.}

     \vskip 0.2cm

     \item {Study all the possible experimental observables that can provide 
            discrimination power between the Signal and Background processes.
            Apply cuts and define the boundaries of these observables to keep 
            Signal efficiency relatively high while try to limit the Background 
            processes as much as possible.
            The cuts and boundaries on these observables determine the data 
            sample, together with all the possible physics hypotheses and their 
            allowed parameter(s) of the physics hypotheses for the data sample.}

     \vskip 0.2cm

     \item {Construct maximum likelihood fit including all hypotheses of
            the physics (Signal and Background) processes contributing to 
            the data sample. 
            The likelihood of an event is parameterized by the sum of 
            probabilities for all relevant Signal and Background hypotheses, 
            with relative weights to be determined by maximizing the likelihood 
            function. The probability of a particular hypothesis is calculated 
            as a product of the PDFs for each of the input experimental 
            observables. Notice the PDFs may depend on some parameters 
            (i.e. the mass of the new particles for Signal and Background) and
            there can be correlations among the various PDFs.}

     \vskip 0.2cm

     \item {Scan all the allowed parameter(s) space and perform unbinned 
            maximum likelihood fit for each set of the parameter(s) of all 
            possible physics hypotheses in the data sample to find the physics 
            hypotheses and their parameter(s) that give best fit to the data.}

     \vskip 0.2cm

     \item {The significances of the favored physics hypotheses and their 
            parameter(s) are determined from how far away statistically they 
            are from all the other possible physics hypotheses and parameter(s).}

  \end{itemize}

\section{Comments on Using Maximum Likelihood Method to Search for New 
         Particles at LHC}

The advantage of this alternative new strategy described above is the high 
efficiency and sensitivity to Signal process. However, they come with a price.

We need large Monte Carlo samples of  Signal and Background processes for 
each set of the allowed parameter(s) in order to obtain the PDF shapes of 
all the physics hypotheses with all allowed parameter(s). The PDF shapes will 
also depend on these parameter(s) and there are in general correlations among 
these observables described by the PDFs.

To determine the cuts and the boundaries of the experimental (discriminative) 
observables also require lots of work. 
In the messy hadron collider environment, we want to keep the Signal process 
as efficient as possible while try to limit the Background processes, the 
possible physics hypotheses and their parameter(s) as much as possible. 
We may face technical difficulties in some searches where the allowed physics
hypotheses and parameters become just too large to handle.
To simplify the possible physics hypotheses and allowed parameter(s) and to 
be conservative, we can even choose the cuts and boundaries on the 
discriminative observables to be very close to the event selection of the 
current cut analysis. The signal efficiency would be similar to the current 
cut analysis, but we get rid of the additional statistical fluctuation 
probability problem.

To scan over the allowed parameter(s) space, handle all the correlations among
all the PDFs and perform unbinned maximum likelihood fit for each set of the 
parameter(s) will certainly be very CPU intensive. 

Needless to say that extensive tests using all possible independent Monte 
Carlo Signal, Background and combined samples will be essential to fully 
understand the fit efficiencies for Signal and Background processes, 
crossfeed among various 
processes, the interpretation of the physics hypotheses and their parameter(s) 
which gives best fit to the test samples. Extensive study is necessary to 
determine the sensitivity of each search. The optimization of the sensitivity 
includes the varying of the cuts and boundaries on the experimental 
(discriminative) observables and repeat these studies to find which set of
cuts and boundaries on the experimental observables can result in the best 
sensitivity for the specific search.

Extreme care needs to be taken when apply this maximum likelihood approach to 
real data, as no simulation can be a perfect representation of the real data. 
An important cross-check of the result is the projection to each discriminate 
observable and check how well the expected physics hypotheses fit the data. 
Through we do not base our discovery or significance of our observed signal on 
a statistically significant peak or excess of events, the projection plot with 
a consistent peak or excess of events in the invariant mass or similar kinematic 
observable is the best verification that we do observe a significant new 
signal~\cite{bigrare,cleopv}.
Unless we can demonstrate that the quality of the fit and projections of the 
data on all experimental observables are all reasonable and they clearly 
indicate the existence of a new particle which is significantly incompatible 
with all the other possible physics hypotheses, we can not claim a discovery 
of a new particle.

The experimental observables, their parameterizations and the correlations 
among various PDFs are all different for each search. The number of unknown 
parameters and their allowed ranges will also vary for each search.
Despite all these differences, the analysis structure, framework and procedures 
are the same for these searches using this new strategy. Therefore, the 
development of the common analysis structure, packages and tools for generic 
searches using this new analysis approach will be extremely useful.

\section{Acknowledgment}

\noindent
The author would like to thank the members of the SMU HEP 
group for their encouragements and useful discussions. 
This work is supported by the U.S. Department of Energy under grant
number DE-FG03-95ER40908 and the supplemental grant to the author
during his research leave of Fall 2003 semester.

\end{document}